# Revealing the Charge Density Wave Proximity Effect in Graphene on 1T-TaS$_2$

**Michael A. Altvater[1+], Sheng-Hsiung Hung[2+], Nikhil Tilak[1], Choong-Jae Won[3], Guohong Li[1], Sang-Wook Cheong[1], Chung-Hou Chung[4,5], Horng-Tay Jeng[2,5*] and Eva Y. Andrei[1*]**

[1] Department of Physics and Astronomy, Rutgers, the State University of New Jersey, 136 Frelinghuysen Rd, Piscataway, New Jersey 08854, USA
[2] Department of Physics, National Tsing-Hua University, 101 Guangfu Road, Hsinchu 30013, Taiwan
[3] Laboratory for Pohang Emergent Materials, Pohang Accelerator Laboratory and Max Plank POSTECH Center for Complex Phase Materials, Pohang University of Science and Technology, Pohang 790-784, Korea
[4] National Chiao Tung University, 1001 Daxue Road, Hsinchu 30010, Taiwan
[5] Physics Division, National Center for Theoretical Sciences, Taipei 10617, Taiwan

*Correspomding authors E-mail: jeng@phys.nthu.edu.tw, eandrei@physics.rutgers.edu,
[+]Equal contributors

Proximity effect is a very powerful approach and has been widely applied to induce electron correlations such as: superconductivity, magnetism and spin-orbit effects at the interface of heterostructure quantum materials. However, proximity induced charge density wave (CDW) state has remained elusive. We report the first observation of a novel proximity induced CDW within a graphene layer that is deposited on 1T-TaS$_2$ crystal. By using scanning tunneling microscopy and spectroscopy to probe the interface of the graphene/1T-TaS$_2$ heterostructure together with theoretical modeling, we show that the interactions between the Dirac-like carriers in graphene and the correlated electrons in 1T-TaS$_2$ induce a periodic charge density modulation within graphene and modify the band structure at the surface of 1T-TaS$_2$, resulting in a 7.5% reduction of its gap size. Our results provide a new platform to manipulate the electron charge correlations in heterostructures.

The isolation and manipulation of atomically thin layered materials provides a ready-made two dimensional electron system [1] whose properties can be tuned by external knobs, such as stress or substrate morphology [2-4], leading to the emergence of novel electronic effects. Distinct from these external knobs, a very powerful approach to manipulate electron correlations of these materials is by contact proximity effect. It is well known that combining materials that host correlated electron phases, such as superconductivity [5] or magnetism [6], with normal metals, gives rise to proximity effects where the correlations persist into the normal metal. These contact proximity effects are a direct consequence of the quantum mechanical properties of electrons in solids; specifically, the non-local nature of electrons. As quantum particles do not have a well-defined position, electronic states cannot abruptly change from one type of ordering to another at the interface of two materials. Consequently, correlated states persist into the normal metal where scattering events begin to destroy the coherence (and vice versa). In the case of 2D materials where scattering is reduced due to their atomically sharp interfaces, proximity-effects are particularly robust allowing correlated states to persist over long distances. The discovery of graphene and other 2D materials, together with the technology enabling 2D heterostructures has led to the observation of strong proximity effects at the atomic limit including proximity induced superconductivity, magnetism and spin-orbit effects [7-15].

The CDW is yet another extensively studied quantum many-body state arising from electron correlations, such as: on-site Coulomb repulsion and short-ranged anti-ferromagnetic spin-exchange interactions. It has been predicted and observed in various correlated insulators and unconventional superconductors [16]. In recent years various CDWs states have been discovered in 2D Transition-metal dichalcogenide (TMDs) [17]]. However, to date inducing a proximity effect between a CDW material and a normal metal has remained elusive as interface effects are highly screened in 3D metals, obscuring measurement, and the CDW state is highly sensitive to defect scattering at the material surface. The ability to observe the charge distribution in graphene directly with sensitive local probes, the use of 2D materials with atomically smooth surfaces, and the ability to stack them [in inert atmosphere] without inducing damage or contamination at the interface allow us to overcome these hurdles.

In this work, we present microscopic evidence of the proximity effect between the CDW material 1T-$TaS_2$, and graphene. Through scanning tunneling microscopy (STM) and spectroscopy (STS), we show that the charge density modulation in 1T-$TaS_2$ persists within the contacted graphene layer. By comparing with first-principles calculations based on density-functional theory (DFT), we identify the global charge transfer between the graphene and 1T-TaS2 surfaces caused by the relative electron negativity and nontrivial local charge transfer that lead to the CDW on graphene. Finally, we demonstrate a model of local charge transfer based on the second order charge transfer process which describes the proximity-induced CDW in graphene.

Samples were fabricated by mechanical exfoliation of graphene and separately $TaS_2$ flakes inside an argon-filled glovebox. The thin $TaS_2$ flake (2.4-24nm or 4-40 layers) was exfoliated from a bulk 1T-$TaS_2$ crystal, which was grown by iodine chemical vapor transport, and transferred onto a passivated $SiO_2$-capped degenerately doped Si wafer. The graphene and 1T-$TaS_2$ flakes were aligned vertically and brought into close contact with micromanipulators under an optical microscope and then heated to promote adhesion. Standard electron beam lithography and electrode deposition (4-5nm Ti/40-50nm Au) were used to make electrical contact to the sample [18]. After removing the PMMA mask the resulting heterostructure was annealed (180-220°C) in hydrogen/argon (10%/90%) to remove polymer residues [19]. STM and STS were performed using a homebuilt STM [20,21] at 78K in high-vacuum <$10^{-5}$ Torr. At this temperature we partially remove the encapsulating graphene layer using the STM tip (mechanically cut Pt/Ir ) resulting in a region that allows us to differentiate the properties of bare and graphene covered 1T-$TaS_2$ (see Supplementary Figure S1). We use an RHK R9 SPM controller for electronic control and data acquisition.

1T-$TaS_2$ is known to exhibit a strongly coupled, commensurate CDW below ~180K involving 13 unit cells where 12 of 13 Ta atoms displace from their high-temperature, equilibrium positions toward the central, 13$^{th}$ Ta atom. This lattice displacement and corresponding charge density modulation repeat periodically within each layer to form a triangular lattice with spacing of $\sqrt{13}$ unit cells of 1T-$TaS_2$. Fig. 1(a) left panel shows the $\sqrt{13} \times \sqrt{13}$ CDW reconstructed 1T-$TaS_2$ supercell. As shown in Fig. S3, the origin of the CDW formation in 1T-$TaS_2$ is the Kohn anomaly in the acoustic branch of the phonon spectrum along the Γ→M direction with the phonon frequency critically suppressed by electron-phonon interaction [22,23], leading to the static displacement of the lattice with the wavevector $\mathbf{Q_{CDW}}$ corresponding to the $\sqrt{13} \times \sqrt{13}$ CDW reconstruction. This soft phonon mode at the Kohn anomaly wavevector $\mathbf{Q_{CDW}}$ consists primarily of longitudinal vibrations of Ta atoms with a minor contribution from transverse vibrations of S atoms relative to the phonon propagating direction $\mathbf{Q_{CDW}}$. The star-shape atomic arrangement involves the local lattice contraction around the center of the David star, in which the bond lengths between Ta ions are shorter in comparison with those between Ta ions outside the star. To compare with STM images, we calculated the partial charge density of the two lowest energy bands as shown in Fig. 1(a) right panel. The electron cloud mainly distributes over the stars with the highest charge density enhancement of ~$5\times10^{-3}$ e/Å$^3$ locating at the center Ta ion of the star.

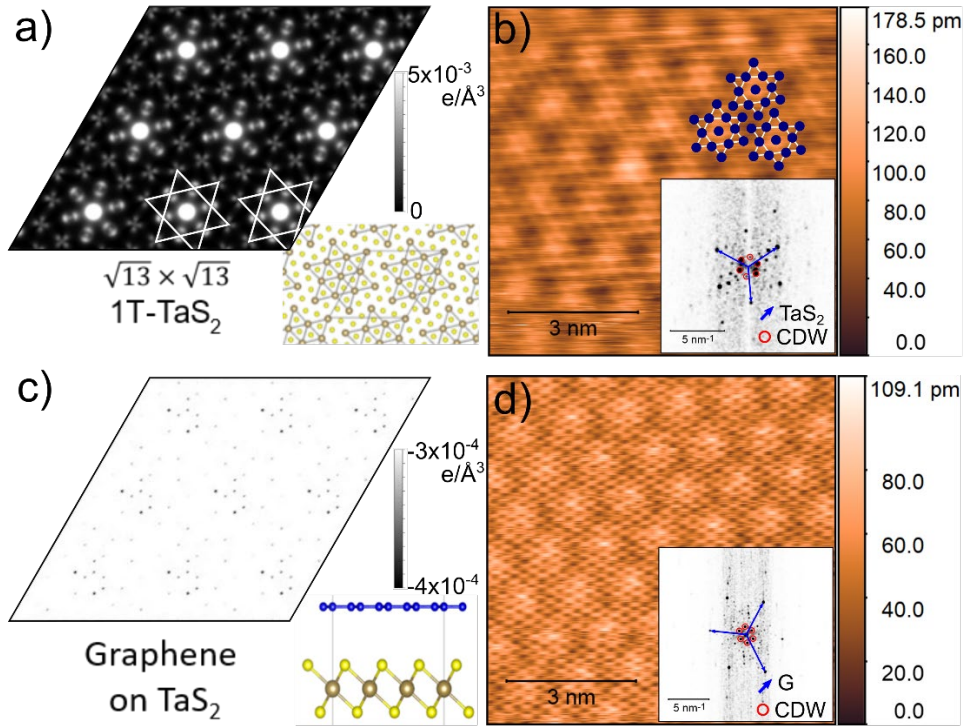

**Figure 1 STM Topography of Bare and Encapsulated 1T-TaS2:** a) Left panel: Ball and stick model of the $\sqrt{13}\times\sqrt{13}$ CDW reconstructed 1T-TaS$_2$ lattice. Right panel: DFT computed charge density of the two lowest energy bands in the commensurate CDW b) STM topography image of the bare surface of 1T-TaS$_2$ shows the CDW charge modulation as well as the atomic lattice of the 1T-TaS$_2$ top layer (Ta positions marked with blue dots) (V$_b$=1.2V, I$_{SP}$=40pA) inset: FFT of topography image shows Bragg peaks from the lattice as well as the CDW modulation. c) Right panel: Ball and stick model of graphene placed on 1T-TaS$_2$ in the commensurate phase. Left panel: DFT computed charge transfer of the graphene layer (the charge density of pristine graphene is subtracted from the charge density of the graphene layer on TaS$_2$) shows a local modulation of doping with the periodicity of the CDW of TaS$_2$. The pattern of the black dots (negative) indicate stronger hole doping in graphene around the David stars. d) STM topography image of graphene placed on 1T-TaS$_2$ shows both the graphene lattice as well as a larger modulation of charge density associated with the CDW of 1T-TaS$_2$ inset: FFT of topography image shows Bragg peaks from the graphene lattice as well as the CDW modulation.

Experimentally, the charge density modulation is evident in the STM topography image of the bare 1T-TaS$_2$ surface, Fig. 1(b). The STM measures both the density of states (DOS) modulation due to the CDW reconstruction and the positions of the sulfur sub-lattice. This can be seen in the fast Fourier transform (FFT) [Fig. 1(b) inset] which shows Bragg peaks associated with the 1T-TaS$_2$ lattice spacing (blue arrows) as well as peaks corresponding to the CDW wavevectors (red circles). The $\sqrt{13}\times\sqrt{13}$ CDW reconstruction is depicted as star-shaped clusters of Ta atoms (blue dots) in Fig. 1(b).

**Figure 2: GGA+U Band Structures of 1T-TaS$_2$ and Graphene on 1T-TaS$_2$** a) GGA+U band structure of $\sqrt{13} \times \sqrt{13}$ CDW reconstructed 1T-TaS$_2$ with U=2.27eV. Hubbard bands, associated with the localized

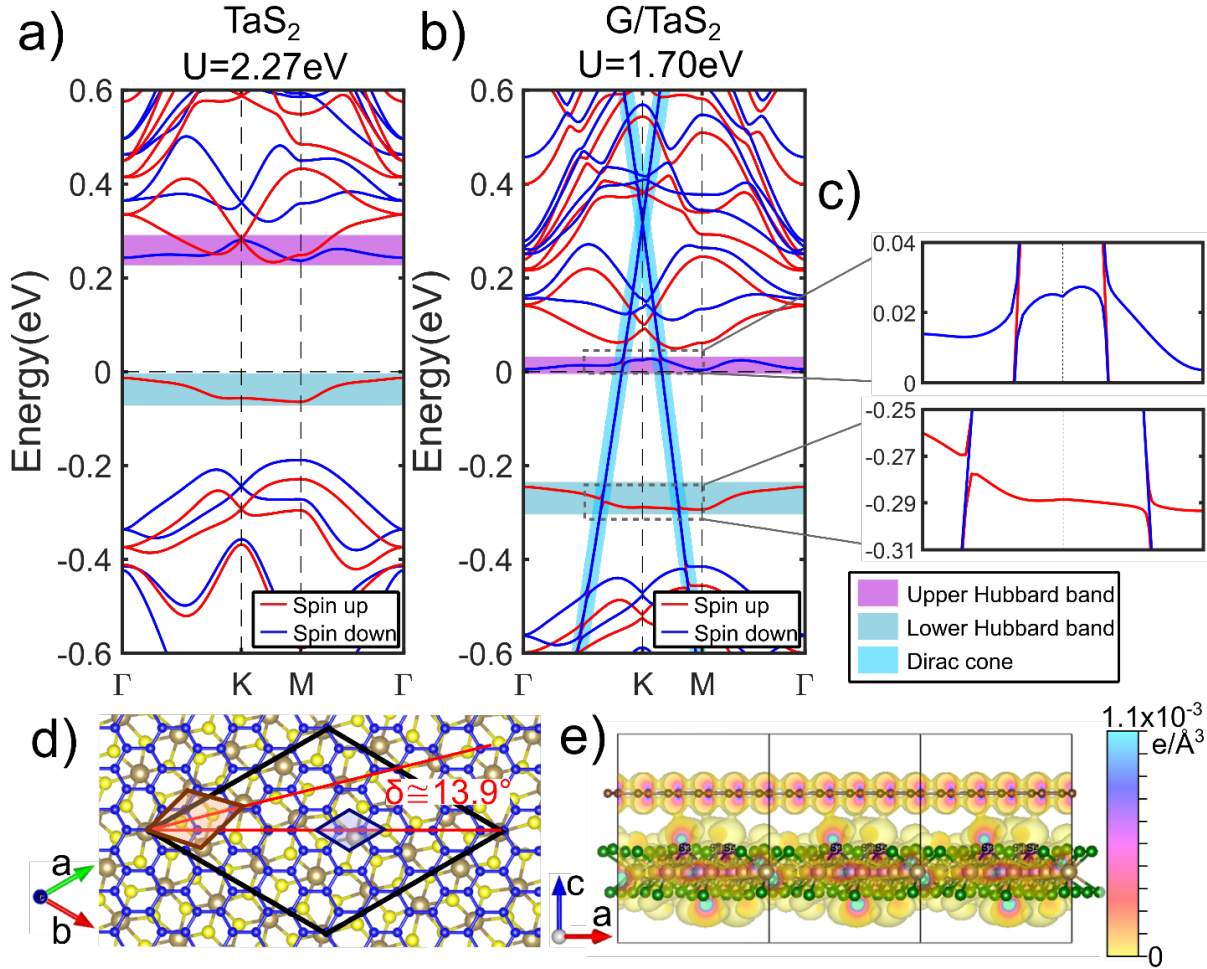

electronic state at the center of the David stars, are highlighted. b) GGA+U band structure of G/TaS$_2$ using a phenomenological value U=1.70eV. Owing to the charge transfer from graphene to 1T-TaS$_2$, the Fermi level (zero energy) moves from the lower Hubbard band to the upper Hubbard band, and the graphene-associated Dirac point at the K-point of the superstructure Brillouin zone is shifted to ~ 0.3eV above the Fermi level (E$_F$) indicating hole doping. The hole distribution [Fig. 1(c)] forms the real space modulation in graphene with the same periodicity as the underlying √13×√13 CDW of 1T-TaS$_2$ through the proximity effect. Hubbard bands and Dirac cone are highlighted for clarity. c) Zoomed in view of the crossing points between the Dirac cone and Upper (top) and Lower (bottom) Hubbard bands. d) Top view of G/TaS$_2$ heterostructure. Blue, brown, and yellow spheres indicate C, Ta, and S atoms, respectively. Black, blue, and brown rhombuses show the 5x5 G/$\sqrt{13} \times \sqrt{13}$ TaS$_2$ supercell, graphene 1x1 unit cell, and 1T-TaS$_2$ 1x1 unit cell, respectively. The graphene and TaS$_2$ layer are twisted by ~13.9° in this CDW phase. e) Side view of G/TaS$_2$ heterostructure overlaid with the charge density map corresponding to the states at the two crossing points of the Dirac cone and Lower Hubbard band. Here, brown, grey, and green spheres represent C, Ta, and S atoms, respectively.

First-principles electronic structure calculations were performed using the projector augmented wave (PAW) approach within the framework of density functional theory (DFT) as implemented in the Vienna ab initio Simulation Package (VASP) [24-27]. The exchange-correlation is described in the Perdew-Burke-Ernzerhof (PBE) form of generalized gradient approximation (GGA).[27,28] Fig. 1(c) right panel shows the side view of the graphene/$TaS_2$ (G/$TaS_2$) heterostructure used in the DFT calculation [The supercell structure is shown in Fig. 2(d)]. By subtracting the calculated charge density of freestanding graphene monolayer from the charge density of G/$TaS_2$, Fig. 1(c) left panel illustrates the 1T-$TaS_2$ induced local doping in the graphene cover layer. From the induced charge density map, we observe a negative (dark) induced charge density of ~0.0004 holes/ Å$^3$ near the centers of the David stars in the 1T-$TaS_2$ layer below (local hole doping). The induced charge density modulation in the graphene layer shows the same periodicity as the CDW in 1T-$TaS_2$ layer with the strongest charge transfer located around the center of the 1T-$TaS_2$ CDW stars with the strongest CDW potential,, demonstrating the correlation between graphene and 1T-$TaS_2$. Indeed, STM topography measurements of the top graphene layer in a G/$TaS_2$ heterostructure also reveal a strong intensity modulation of graphene's honeycomb lattice with a period that matches that of the CDW in 1T-$TaS_2$ [see Fig. 1(d)]. The two-dimensional periodic modulation is evident in the fast Fourier transform (FFT) of the topography image [Fig. 1(d) inset]. As detailed below, the charge transfer in graphene layer provides clear evidence of the CDW proximity effect.

To investigate the charge transfer between the two materials at the interface, we calculate the band structure of bare 1T-$TaS_2$ using the lattice-relaxed $\sqrt{13} \times \sqrt{13}$ CDW supercell as well as the G/$TaS_2$ heterostructure by placing 5x5 unit cells of graphene on top. The top and side views of the G/$TaS_2$ lattice model are depicted in Figs. 2(d) and (e), respectively. To take the strong correlation of Ta d-electrons into consideration, we perform generalized-gradient approximation plus on-site U (GGA+U) calculations with U=2.27eV for bare, monolayer 1T-$TaS_2$ in accordance with previous DFT calculations [29,30] as shown in Fig. 2(a). The CDW-induced isolated half-filled spin-degenerate flat band at $E_F$ [see Fig. S4] splits into occupied spin up lower Hubbard band (LHB) and empty spin down upper Hubbard band (UHB) with a Mott gap of ~0.25eV in between. The on-site Hubbard U significantly enhances the Mott gap from ~0.12eV [see Fig. S5] to ~0.25eV [Fig. 2(a)] so that the UHB touches the lowest conduction bands, while experiments observe a localized Hubbard band more separated from the conduction band. This discrepancy implies the on-site Hubbard U of 2.27eV might be somewhat overestimated. Possible reasons are discussed below.

In the band structure of the G/$TaS_2$ heterostructure, both the 1T-$TaS_2$ Hubbard bands and the graphene Dirac cone are preserved [see Fig. 2(b)]. Compared to bare 1T-$TaS_2$, the Fermi level moves from the LHB up to the UHB due to the charge transfer from graphene to 1t-$TaS_2$. This is also accompanied with an energy shift of the graphene Dirac point from the Femi level $E_F$ up to ~0.3eV above $E_F$ at the K-point of the superstructure Brillouin zone [Fig. S3]. A careful comparison of the Dirac cone of G/$TaS_2$ with that of bare graphene leads us to conclude that the Fermi velocity is not changed by the proximity to 1T-$TaS_2$. The highly dispersive pz-orbital derived Dirac cone intersects the CDW and Mott gaps and crosses both the $d_{z^2}$-orbital derived narrow LHB and UHB. They indeed interact with each other, though in a gentle manner. Four small gaps of order 10meV emerge at the four crossing points as shown in the insets of Fig. 2(c). These small gaps which originate from the weak interlayer couplings indicate finite hybridization between the Dirac and Hubbard states, giving rise to the proximity effect between graphene and 1T-$TaS_2$ as illustrated in Fig. 2(e).

Graphene is known to be a wide bandwidth metal with high-mobility itinerant carriers. With graphene on top of 1T-TaS$_2$, there exists nontrivial charge transfer from graphene to 1T-TaS$_2$, as shown in the LDSA+U band structure in Fig. 2(b). These highly mobile electrons from graphene somewhat suppress the localized picture of the narrow Hubbard d-bands in 1T-TaS$_2$ and thus reduce the on-site U value of Ta. We have compared our band structures for different U values with our STS results, suggesting that the graphene layer screens the Coulomb interaction in 1T-TaS$_2$ and the Hubbard U of Ta is lowered by ~0.5eV due to the itinerant electrons from graphene. Therefore, we show in Fig. 2(c) the G/TaS$_2$ band structure with a phenomenological value U=1.70eV, which reasonably reproduces the trend observed in our STS measurements (below).

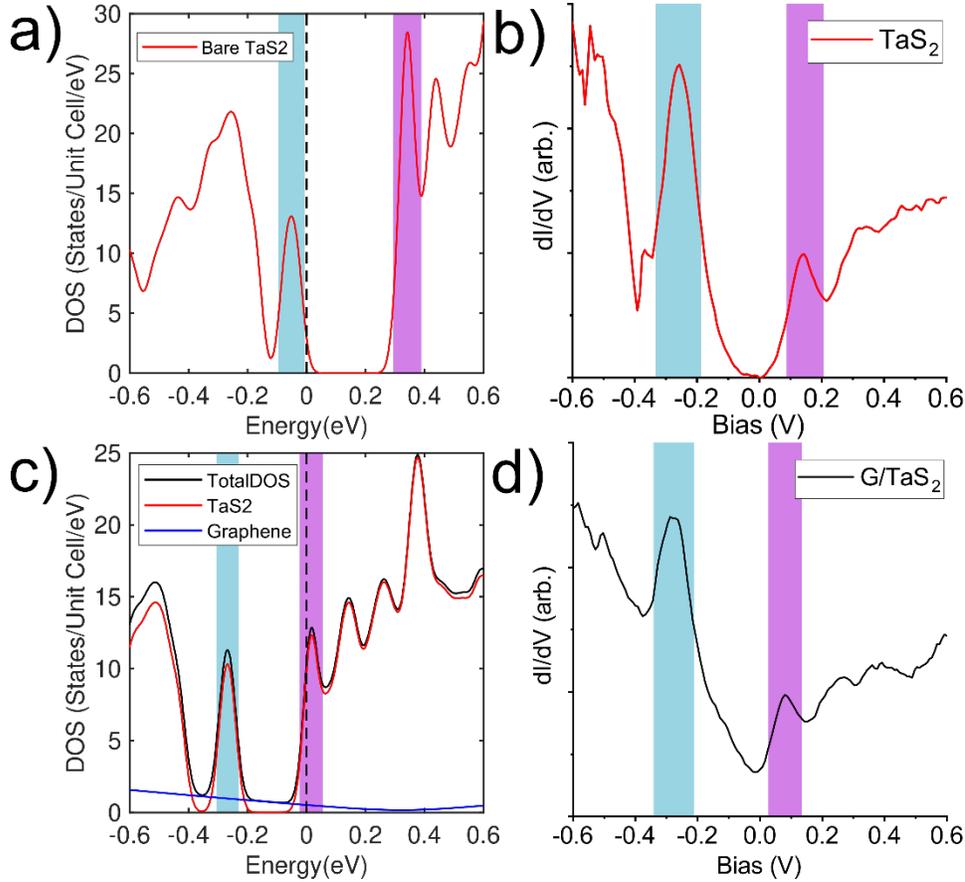

**Figure 3: Density of States of Bare and Graphene Encapsulated 1T-TaS$_2$** a) GGA+U calculated DOS of bare 1T-TaS$_2$ (U=2.27eV) b) Measured tunneling spectroscopy (STS) of the bare 1T-TaS$_2$ surface shows a gap at zero bias with peaks associated with upper and lower Hubbard bands (highlighted with violet and blue, respectively) in agreement with the calculated result (a). c) GGA+U calculated DOS of the G/TaS$_2$ heterostructure showing features resembling the Hubbard bands, conduction band, and valence band of 1T-TaS$_2$ with a reduced Mott-Hubbard interaction strength (U=1.70eV). The projected DOS (blue line) between the Hubbard-like bands originate from the Dirac cone states within the graphene layer d) Measured STS of G/TaS$_2$ qualitatively agrees with the calculated result in (c), displaying Hubbard-like peaks, with a gap size that is reduced with respect to the Mott gap observed in bare TaS$_2$, and mid-gap states which we associate with the graphene layer. Spectra in (b),(d) are taken at a set point of [V$_b$=1.2V, I$_{SP}$=80pA] with AC modulation $V_{AC} = 8mV$. Lower and upper Hubbard peaks are highlighted in blue and violet, respectively.

Next, we compare the calculated DOS with measured STS (proportional to the local DOS) on both the bare 1T-TaS$_2$ [Fig. 3(a),(b)] and graphene covered regions (Fig. 3c,d). GGA+U DOS of bare 1T-TaS$_2$ shown in Fig. 3(a) reflect the key features of the 1T-TaS$_2$ band structure in Fig. 2(a). The lower and upper Hubbard bands located respectively below E$_F$ and ~0.35 eV above E$_F$ with a ~0.35 eV Mott gap in between. The dips below LHB and above UHB indicate the scale of the CDW gap [Fig. S6]. These main features are consistent with those observed in our STS measurement [shown in Fig. 3(b)] which also show the expected Mott gap at the Fermi level flanked by lower and upper Hubbard peaks at -256mV and 144mV, respectively. Additionally, the two large dI/dV dips observed at -378mV and 218 mV are associated with the known CDW gap.

With graphene on top of 1T-TaS$_2$, the GGA+U DOS plotted in Fig. 3(c) also show UHB, LHB, Mott gap, and CDW gap similar to the DOS of bare 1T-TaS$_2$. The main differences from the bare 1T-TaS$_2$ case are i) the relative energies of the Fermi level, UHB, LHB, and Dirac point; ii) the linearly dispersing states within the Mott gap. Owing to the relatively stronger electron negativity of TaS$_2$, there exist notable charge transfer from graphene to 1T-TaS$_2$. Consequently, the Fermi level moves from the LHB top to the UHB bottom, meanwhile the graphene Dirac point shifts to a higher energy about 0.3eV above E$_F$. On the other hand, the mid-gap linear band comes from the Dirac cone states of the graphene cover layer. Because of the energy shift of the Dirac point to ~0.3eV above E$_F$, the V-shape DOS thus moves to ~0.3eV, leaving a linear-like band within the Mott gap below E$_F$.

The main features given from GGA+U are in good agreement with our STS results shown in Fig. 3(d). Two peaks near the Fermi level are labelled as lower and upper Hubbard bands (measured at -280mV and 88mV, respectively) with a V shaped DOS in between. Unlike the case of pristine, undoped graphene on an insulating substrate, here the DOS does not vanish at the Fermi level. The negative slope of the measured STS in between the Hubbard bands suggests the Dirac point in graphene has shifted to higher energies, indicative of the hole doping. We note that the GGA+U calculation of the DOS [Fig. 3(c)] is averaged over the entire heterostructure whereas the measured STS favors the region closest to the tip [Fig. 3(d)]. The preferential tunneling into the graphene top layer results in a larger mid-gap STS intensity (compared to Hubbard peak heights) than given by the calculation. Further modelling is needed to reproduce the relative spectral weights due to different orbitals within the vdW heterostructure as measured by an STM tip above the surface.

Comparing STS measurements of bare 1T-TaS$_2$ [Fig. 3(b)] and G/TaS$_2$ [Fig. 3(d)], we can see that the addition of the graphene layer induces both a shift of the Hubbard peaks as well as a reduction of the gap between them by approximately 7.5%. We interpret the reduction of the Mott gap size as being due to screening of the Coulomb interaction near the surface of 1T-TaS$_2$ by itinerant electrons in graphene, reducing U and the separation between Hubbard peaks. As mentioned above, we adopt a smaller on-site U=1.70eV in the calculation of the G/TaS$_2$ case to consider this screening effect observed in STS.

We note that taking a unit cell consisting of two layers of 1T-TaS$_2$ results in dimerization between layers, as suggested by recent theoretical and experimental works [31-34]. In this case, 1T-TaS$_2$ is no longer best described as a two-dimensional Mott insulator with magnetic ordering, but rather a band insulator, as the formation of interlayer dimer singlets suppresses magnetism. We show in Supplemental Material [Fig. S7] that the main features of adding a graphene layer on top of one or two layers of 1T-TaS$_2$ are qualitatively similar to the single layer case, giving rise to a shift of E$_F$, reducing the gap size, and featuring charge transfer with the graphene layer. Our experimental technique is not sensitive to the spin-polarization; thus, we cannot identify the magnetic ground state of 1T-TaS$_2$ in this study. However, our calculations suggest that if a magnetic moment is present in 1T-TaS$_2$, the addition of the graphene layer will not suppress the moment at the surface. Therefore, measurements using graphene encapsulation and spin-polarized STM tips might be able to identify the magnetic properties of the 1T-TaS$_2$ surface in future measurements.

As mentioned above, the charge density map in Fig. 1(c) right panel shows the hole density in graphene induced by 1T-TaS$_2$, which corresponds to the local charge transfer from graphene to TaS$_2$. The missing electron density in graphene shows a clear correlation with the CDW pattern in 1T-TaS$_2$, indicating a proximity-induced CDW in graphene. Figure 2(e) demonstrates the real-space interlayer hybridization arising from the state around the Dirac-LHB crossing points in reciprocal space. The interlayer charge density overlap indicates that the graphene and 1T-TaS$_2$ layers are intimately connected by tunneling electrons, resulting in a novel proximity effect induced by charge transfer as discussed below.

The simplified Hamiltonian of the G/TaS$_2$ bilayer system is given by:

$$H = H_d + H_c + H_t,$$

$$H_d = \sum_{\langle i,j \rangle,\sigma} -t_{ij}^d d_{i,\sigma}^\dagger d_{j,\sigma} + h.c. - \sum_{\langle i',j' \rangle,\sigma} (\Delta_d^{CDW}(i',j'))^* d_{i',\sigma}^\dagger d_{j',\sigma} + h.c. + \sum_{\langle i',j' \rangle} |\Delta_d^{CDW}(i',j')|^2,$$

$$H_c = \sum_{\langle i,j \rangle,\sigma} -t_{ij}^c c_{i,\sigma}^\dagger c_{j,\sigma} + h.c. = \sum_{k,\sigma} (\epsilon_k - \mu) c_{k,\sigma}^\dagger c_{k,\sigma},$$

$$H_t = -t \sum_{i,\sigma} c_{i,\sigma}^\dagger d_{i,\sigma} + h.c.,$$

where $H_d$ ($H_c$) stands for the simplified Hamiltonian of the 1T-TaS$_2$ (graphene) layer, respectively, and $H_t$ describes a weak charge transfer (hopping) term between these two layers (we neglect a small mismatch in spatial locations between the nearest-neighbor sites of the corresponding layer). The insulating 1T-TaS$_2$ layer shows CDW order with the order parameter $\Delta_d^{CDW}(i',j') \equiv \sum_\sigma \langle d_{i',\sigma}^\dagger d_{j',\sigma} \rangle$ where $i',j'$ are sites within the CDW unit cell. The graphene layer has the linear Dirac dispersion: $(\epsilon_k - \mu) \approx \hbar v_F |k - k_F|$. Here, $i,j$ refer to the nearest-neighbor sites of the corresponding lattices, and $t_{ij}^{d(c)}$ refers to the nearest-neighbor tight-binding hoping terms on the 1T-TaS$_2$ (graphene) layer, respectively.

Via the second order perturbation in the $H_t$ term of the CDW unit cell, the following exchange term $H_t^{(2)}$ is generated:

$$H_t^{(2)} = t^2 \sum_{\langle i',j' \rangle,\sigma,\sigma'} c_{i',\sigma}^\dagger d_{i',\sigma} d_{j',\sigma'}^\dagger c_{j',\sigma'} + h.c.$$

A simple mean-field decoupling of $H_t^{(2)}$ in terms of $\Delta_d^{CDW}(i',j')$ (considering only $\sigma = \sigma'$ and assuming spin-isotropic CDW order $\langle d_{j',\uparrow}^\dagger d_{j',\uparrow}\rangle = \langle d_{j',\downarrow}^\dagger d_{j',\downarrow}\rangle$) gives $H_t^{(2)} \to H_{t^2}^{MF}$ with:

$$H_{t^2}^{MF} \approx -t^2/2 \sum_{\langle i',j'\rangle,\sigma} (\Delta_d^{CDW}(i',j'))^* c_{i',\sigma}^\dagger c_{j',\sigma} - t^2/2 \sum_{\langle i',j'\rangle,\sigma} (\Delta_d^{CDW}(i',j'))^* \langle c_{i',\sigma}^\dagger c_{j',\sigma}\rangle + h.c.,$$

where the mean-field decoupling term ($\langle c_{i,\sigma}^\dagger c_{j',\sigma}\rangle d_{i',\sigma}^\dagger d_{j',\sigma}$) in $H_t^{(2)}$ is neglected since we expect $\left|\langle c_{i',\sigma}^\dagger c_{j',\sigma}\rangle\right| \ll \left|\langle d_{i',\sigma}^\dagger d_{j',\sigma}\rangle\right|$. The CDW proximity effect is manifested in $H_{t^2}^{MF}$ as a weak CDW order $\sum_\sigma \langle c_{i',\sigma}^\dagger c_{j',\sigma}\rangle$ is induced on the graphene layer by the second order charge transfer between the two layers with the following identification:

$$\Delta_c^{CDW}(i',j') \equiv -1/2 \sum_\sigma \langle c_{i',\sigma}^\dagger c_{j',\sigma}\rangle^* = -t^2/2 \left(\Delta_d^{CDW}(i',j')\right)^*,$$

or equivalently, $\sum_\sigma \langle c_{i',\sigma}^\dagger c_{j',\sigma}\rangle = t^2 \Delta_d^{CDW}(i',j')$. Via the above identification, the Hamiltonian $H_{t^2}^{MF}$ can be expressed as:

$$H_{t^2}^{MF} = \sum_{\langle i',j'\rangle,\sigma} \Delta_c^{CDW}(i',j') c_{i',\sigma}^\dagger c_{j',\sigma} + h.c. + 2|\Delta_c^{CDW}(i',j')|^2,$$

which leads to $\Delta_c^{CDW}(i',j') = -1/2 \sum_\sigma \langle c_{i',\sigma}^\dagger c_{j',\sigma}\rangle^*$ identified above via minimizing the free energy associated with $H_{t^2}^{MF}$ with respect to $\Delta_c^{CDW}(i',j')$. Note that from above derivations, we indeed find that $\left|\langle c_{i',\sigma}^\dagger c_{j',\sigma}\rangle\right| \sim t^2 \left|\langle d_{i',\sigma}^\dagger d_{j',\sigma}\rangle\right| \ll \left|\langle d_{i',\sigma}^\dagger d_{j',\sigma}\rangle\right|$, as expected. Note also that the CDW order parameters induced on graphene layer shows the opposite sign with respect to that on 1T-TaS$_2$ layer, consistent with the hole-like (particle-like) CDW intensity on graphene (1T-TaS$_2$) layer obtained from DFT calculations, respectively. We emphasize here that the above mechanism based on charge transfer is distinct from all the previously realized proximity effects, including superconducting, magnetic and spin-orbit proximity effects.

In summary, we demonstrate, by STM, STS measurements and DFT calculations, the existence of a novel proximity-induced CDW in graphene resulting from its contact with a 1T-TaS$_2$ crystal. Meanwhile, the graphene cover layer profoundly affects the insulating phase in 1T-TaS2 by providing mid-gap carriers which screen the Mott-Hubbard interaction term, U, and reduces the size of the Mott gap. The periodic modulation of interlayer coupling, caused by the CDW reconstruction within 1T-TaS$_2$, acts as an external potential applied to graphene similar to that induced by the Moiré pattern in twisted bilayer graphene. This proximity induced CDW is characterized by a charge density modulation that is out-of-phase with that of the host material even though they have the same wavelength. Our model captures the CDW proximity effect arising from the short-range exchange interaction via second-order local electron hoping process. This observation opens interesting questions regarding this novel proximity-induced CDW, including the possibility of sliding, pinning, and phonon softening of the CDW within the graphene layer. Our results therefore provide a new platform to manipulate the electron charge correlations in heterostructures. Further measurements including STM/STS of graphene at the edge of 1T-TaS$_2$, Raman spectroscopy, and electronic transport measurements of the heterostructure will further elucidate the effects proposed in this work.

# Acknowledgements

MAA was supported by the National Science Foundation grant EFRI 1433307; EYA and NT acknowledge support from the Department of Energy grant DOE-FG02-99ER45742 and the Gordon and Betty Moore Foundation EPiQS initiative grant GBMF9453; GL was supported by Rutgers University; CJW was supported by Max Planck POSTECH/KOREA Research Initiative Program, Grant 2011-0031558; SWC was supported by the Betty Moore Foundation's EPiQS grant GBMF6402 and Rutgers University, C.-H. C. was supported by MOST (Grant NO.: 107-2112-M-009-010-MY3, 110-2112-M-A49-018-MY3) and the NCTS of Taiwan, R.O.C., J.H.T acknowledges support fromthe Ministry of Science and Technology, Taiwan under grant: MOST 109-2112-M-007 -034 -MY3, and from NCHC, CINC-NTU, AS-iMATE-109-13, and CQT-NTHU-MOE, Taiwan.